\documentclass[12pt]{article}

\usepackage{amsfonts}
\usepackage{amssymb}
\usepackage[dvips]{graphicx}

\newcommand{\be}{\begin{equation}}
\newcommand{\ee}{\end{equation}}
\newcommand{\ba}{\begin{eqnarray}}
\newcommand{\ea}{\end{eqnarray}}
\newcommand{\eps}{\epsilon}

\newcommand{\lie}{{\cal L}}
\newcommand{\Lie}{\lie}
\newcommand{\ag}{\alpha}
\newcommand{\bg}{\beta}
\newcommand{\cg}{\gamma}
\newcommand{\dg}{\delta}
\newcommand{\eg}{\rho}

\begin{document}
\title{Mechanics of multidimensional isolated horizons}
\author{Miko\l{}aj\ Korzy\'nski${}^{1}$,
Jerzy\ Lewandowski${}^{2,1,3}$, Tomasz Paw\l{}owski$^{1,2}$}
\date{ }

\maketitle
{\it 1. Instytut  Fizyki Teoretycznej, Uniwersytet
Warszawski, ul. Ho\.{z}a 69, 00-681 Warsaw, Poland\\
2. Center for Gravitational Physics and Geometry, Physics
Department, 104 Davey, Penn State, University
Park, PA 16802, USA\\
3. Max--Planck--Institut f\"ur Gravitationsphysik, Albert--Einstein--Institut,
 14476 Golm, Germany}

\begin{abstract}


Recently a multidimensional generalization of the isolated horizon
framework has been proposed
\cite{LP}. Therein the geometric description was easily
generalized to higher dimensions and the structure of the
constraints induced by the Einstein equations was analyzed. In
particular, the geometric version of the zeroth law of the black
hole thermodynamics was proved. In
this work we show how the IH mechanics can be formulated in a
dimension--independent fashion and derive the first law of BH thermodynamics
for
arbitrarily dimensional IH. We also propose a definition of energy
for non--rotating horizons.

\end{abstract}
\medskip

\section{Introduction}

Four--dimensional isolated horizons proved to be a useful tool
for studying black hole mechanics, thermodynamics and even quantum theory
\cite{ashtekar-baez-krasnov}, \cite{AK}.
Being quasi--local they are also useful for numerical relativity
(gravitational waves investigation, black hole merger studies).
It is therefore interesting to investigate whether the notion of IH exists
in higher dimensions and check if their physical properties are similar.


The Hamiltonian formulation of General Relativity combined with the
IH framework made it possible to formulate
the first law of black hole thermodynamics
for both rotating and non--rotating black holes \cite{AFK}, \cite{ABL-mech},
\cite{CJK2}.
Consider space-times with axially symmetric IH's (the space-times themselves
don't have to be symmetric at all).
The first law arises naturally as one investigates the transformations
of the phase space given by  flows of vector fields (time translation).
One may ask about the conditions for such transformations to be generated by a
Hamiltonian. It turns out (see \cite{AFK}) that the only 
condition
constrains the value of vector field on the horizon. The condition implies that
the generating Hamiltonian must be a function of the horizon area and
angular momentum solely -- but does not constrain this function in any way.
 One can, however, fix the Hamiltonian function
(and therefore the IH e\-nergy value) by requiring that it agrees with
the ADM mass in the case of stationary, asymptotically flat solution. Note that
due to the Kerr solution uniqueness theorems such fixing is consistent:
if the solution is stationary and asymptotically flat, it must be
a Kerr metric
and the ADM mass must depend on $A$ and $J$ in the Kerr--like manner.

This work deals with $N+2$--dimensional ($N > 2$) generalization of
rotating isolated horizons \cite{LP}. We first check that the Hamiltonian
formalism leads to the same conditions on the Hamiltonian function
as in the 4D case. This result is
valid for any rotating (axially symmetric) weakly isolated horizon
in any dimension.
In order to define the \emph{energy}
however, we restrict ourselves to non--rotating ones.
This is due to the fact that the general uniqueness theorem for the axially
symmetric space-times
fails in higher dimensions \cite{EmparanReall}. One cannot assign safely the energy function using
some family of solutions analogous to Kerr solutions in 4D since there
exist other families with different ADM mass for given horizon area
and angular momentum \cite{MyersPerry}, \cite{EmparanReall}.
One would have to argue somehow why the choice
of one family of solutions is more physical than the other.

It is true however that the conditions for the topology of the horizon to be
a sphere and the existence of \emph{two} axes of symmetry
are strong enough
to prove the uniqueness of solution in 5 dimensions \cite{MI}.

Nevertheless, only in the static, \emph{non--rotating} case
there exist general uniqueness theorems for arbitrary dimension and
for $\sigma$--model, vacuum and
charged black holes
\cite{Rogatko1}, \cite{Rogatko2}, \cite{Rogatko3}.
Since we deal with the vacuum case, we will
assume the energy dependence on the horizon area like in the generalized
Schwarzschild case.

\section{Weakly isolated horizons}
In this section we recall the definition of non-expanding horizons
(NEH) \cite{LP}, spell out the definition of weekly isolated
horizons (WIH) and discuss those of their properties 
which will prove relevant in the next
sections.

The following convention of indeces will be adopted:
\begin{itemize}
\item Greek indices will
be used for
objects defined on the whole $N+2$--dimensional tangent space of $M$,
\item
small Latin letters for objects defined on or contained in the
$N+1$--dimensional subspace (tangent to the horizon) and
\item capital Latin letters for the $N$--dimensional subspace (tangent to a cut).
\end{itemize}

The symbol `$d$' will stand for the exterior derivative for any
manifold.

The abstract index notation will be used whenever convenient.

\subsection{Non-expanding horizons}
 Let $\Delta$ be a $N+1$--dimensional null surface in a
$N+2$--dimensional space-time $M$ equipped with a metric tensor
field of the signature $(-,+,...,+)$ which satisfies Einstein
equations with or without cosmological constant,\footnote{In the
main part concerning the first law we will restrict ourselves to the
vacuum case without cosmological constant.} and let $l$ denote a
non-vanishing normal vector. If the expansion of $l$ vanishes
everywhere on $\Delta$, then this property is independent of
the choice of $l$ and we call $\Delta$ a \emph{non--expanding null
surface}. This assumption, combined with a mild energy condition
$T_{\ag\bg} l^\ag l^\bg \geq 0$, implies a restriction \cite{LP}
on the Ricci tensor:
\be
 R_{\ag\bg}\,l^\ag\,l^\bg = 0.
\ee
It also implies that the metric tensor $q$ induced on $\Delta$ (degenerate)
metric is Lie dragged by any null vector field tangent to $\Delta$
\begin{equation}\label{Llq}
\Lie_l q\ = 0.
\end{equation}
It follows that there exists on $\Delta$ a differential 1--form
$\omega{(l)}$ called the \emph{rotation potential}, such that at
every $x\in\Delta$ and for every $X\in T_x\Delta$,
\be\label{rotationpot}
 X^c \, \nabla_c\, l = X^c \, \omega^{(l)}_c l\ee
where $\nabla$ is the covariant derivative.
Moreover,
a non--expanding surface $\Delta$ is called a \emph{non--expanding horizon}
 (NEH) if
there exists an embedding
\be   \hat\Delta \times [0,\,1] \mapsto M \ee
where $[0,\,1]\subset \mathbb{R}$ stands for the interval, such
that:
\begin{enumerate}
\item $\Delta$ is the image,
\item $\hat \Delta$ is an $N$--dimensional compact, connected manifold,
\item for every maximal null curve in $\Delta$ there exists $\hat x \in \hat
\Delta$ such that the curve is the image of $\{\hat x\} \times
[0,\,1]$.
\end{enumerate}

It follows, that the manifold $\hat{\Delta}$ is the space of the
null geodesics generating $\Delta$ and there is a natural
projection
\be
\Pi:\Delta\rightarrow\hat{\Delta}.
\ee

 Roughly speaking, the definition of a non-expanding horizon
 amounts to an
extra condition on the topology of a non-expanding null
surface\footnote{Note that no geodesic completeness assumption has
been made. Therefore a NEH is not assumed to be extending to past
or future infinity. Physically it means that a NEH can be formed
some time in the past as  well as destroyed or distorted some time
in the future}.

\subsection{The definition of weakly isolated horizons}
A \emph{weakly isolated horizon} is a pair: a NEH $\Delta$
equipped with a class $[l]$ of non-vanishing vector fileds, normal to $\Delta$
(\emph{i.e.} null and tangent)  such that:

\begin{itemize}
\item{} $(i)$ for every $l,\,{l'} \in [l]$ there is a positive constant
$b$ such that
\begin{equation}\label{bl}
{l'}\ =\ bl,
\end{equation}
\item{}$(ii)$ the rotation 1--form potential $\omega^{(l)}$ is Lie dragged by
$l$
\be\label{Llomega} \lie_l \omega^{(l)} = 0. \ee
\end{itemize}

The class $[l]$ will be often referred to as the WIH flow. Note that
the rotation one form potential (\ref{rotationpot}) is insensitive
to the constant re-scalings (\ref{bl}). Indeed, for every function
$b:\Delta\rightarrow \mathbb{R}$ and a null, nowhere vanishing
vector field $l$ tangent to a non-expanding horizon, we have
\begin{equation}
\omega^{(bl)}\ =\ \omega^{(l)} + d\ln b.
\end{equation}
Therefore we will skip from now on the suffix $(l)$, whenever a
WIH is given.

We will summarize several basic facts concerning NEH's and WIH's.
For proofs and further explanations see \cite{LP} and also the
four dimensional case results  \cite{ABL-ge}, \cite{ABL-mech}.

To begin with, note that  the integral curves of a vector field
$l$ normal to NEH $\Delta$ are null and geodesic in the sense that
\be l^\bg\,\nabla_\bg l = \kappa^{(l)}\,l \, .\ee
The function-coefficient $\kappa^{(l)}$ is called the
\emph{surface gravity}. Under a bit stronger energy condition that
$T_{\ag\bg} l^\bg$ is causal and future--pointing one may also
prove the \emph{zeroth law of black hole thermodynamics} which
states that
\be d\kappa^{(l)} = {\cal L}_l \omega^{(l)}. \ee

Using the zeroth law it is easy to prove that every non-expanding
horizon $\Delta$ admits a large class of null vector fields $l$,
each of which defines a distinct weakly isolated horizon
\cite{LP}.

 In particular, it follows that the surface gravity is
necessarily constant for every  WIH. If $\kappa^{(l)} = 0$, the
WIH is called \emph{extremal}. Given a WIH there exists a freedom
of rescaling the vector $l^\ag$ by a constant, positive factor
accompanied by the same rescaling of $\kappa^{(l)}$. Therefore in
the extremal WIH case $\kappa^{(l)}$ is determined as $0$.
Otherwise, in the non-extremal case, its exact value depends on the
choice of $l^\ag\in[l]$, while its sign is determined.

\subsection{Good cuts foliation}
Let $(\Delta, [l])$ be a non-extremal weakly isolated horizon.
There exists a natural foliation \cite{ABL-ge}, \cite{LP} of
$\Delta$ distinguished by the geometry of $(\Delta,[l])$. It is
defined as follows: let $\tilde{\Delta}\subset \Delta$ be any leaf
of the foliation. Then, the pullback $\tilde{\omega}_\ag$ of the
rotation 1--form potential onto $\tilde{\Delta}$ is divergence
free,
\begin{equation}
\tilde{q}^{AB}\tilde{D}_A \tilde{\omega}_B\ =\ 0,
\end{equation}
where $\tilde{q}$ is the metric tensor induced on $\tilde{\Delta}$
and $\tilde{D}$ is the corresponding torsion free covariant
derivative. If we assume that the leaves are global cross-sections
of the maximal analytic extension of $\Delta$, then the foliation
formed by them is unique \cite{LP,ABL-ge}. It is called a {\it
good cuts foliation} of a WIH $(\Delta,[l])$.

\subsection{Symmetries and symmetric WIH's}\label{Sec:sym}
We consider non-extremal WIH's $(\Delta,[l])$ here, of induced
metric tensor $q$ and the rotation 1--form potential $\omega$.

 A vector field $X$ tangent to a WIH $(\Delta,[l])$
 is called a  {\it symmetry generator}  whenever it is
 true  that
 \begin{equation}\label{sym}
 \Lie_X l\ =\ al,\ \ \ \Lie_X q\ =\ 0,\ \ \ \Lie_X
 \omega\ =\ 0.
 \end{equation}

For example a vector field of the form $fl$, where $f$ is a
function and $l\in[l]$, is a symmetry generator if and only if
$f=const$. Indeed,
\begin{equation}
\Lie_{fl} \omega\ =\ f\Lie_f\omega + \kappa^{(l)} d f\ =\
\kappa^{(l)}df.
\end{equation}
 Hence, {\it every} WIH admits the null symmetry generators
 $l\in [l]$.

Suppose now  a given non-extremal WIH $(\Delta,[l])$ admits  
another
symmetry generator, a vector field $X$ which is not everywhere null on
$\Delta$. It follows from the first equality in (\ref{sym}) that
the projection $\Pi_*X$ is a well defined vector field on the base
manifold $\hat{\Delta}$. Due to the second equality in (\ref{sym})
we conclude that
\begin{equation}
\Lie_{\Pi_*X}\hat{q}\ =\ 0,
\end{equation}
hence the vector field $\Pi_*X$ is a Killing vector of the metric
tensor defined on the base manifold $\hat{\Delta}$. This shows
that a generic WIH does not admit non-null symmetries. We now use
the good cuts foliation of $\Delta$ to lift $\Pi_*X$ to a vector
field $\tilde{X}^a$ defined on $\Delta$, tangent to the good cuts
at every point, and such that
\begin{equation}
\Pi_*\tilde{X}\ =\ \Pi_*X.
\end{equation}
Therefore,
\begin{equation}
X\ =\ fl\ +\ \tilde{X},
\end{equation}
where the factor $f$ is a function defined on $\Delta$.

We claim that again
\begin{equation}\label{df}
df\ =\ 0.
\end{equation}
To see this, note first that a stronger condition than the first
equation in (\ref{sym}) is necessarily true, namely
\begin{equation}
\lie_Xl\ =\ 0.
\end{equation}
Indeed,
\begin{equation}
0\ = \lie_X \kappa\ =\ \lie_X(l^a\omega_a)\ =\ a\kappa^{(l)},
\end{equation}
due to the last equality in (\ref{sym}). Secondly, let us exercise
the invariant cha\-racter of the good cuts foliation. Introduce a
function $v:\Delta\rightarrow\mathbb{R}$ constant on each leaf of
the good cut foliation, and such that $l^a D_a v=1$. The
derivative $dv$ is uniquely defined by $l$ and the foliation, and
both are preserved by the local flow of $X$, therefore the
derivative is necessarily preserved by the flow of $X$.  Hence
\begin{equation}\label{Xdv}
0\ =\ \Lie_Xdv\ =\ df.
\end{equation}

Finally, it follows that the vector field
\begin{equation}
\tilde{X}\ =\ X\ -\ fl
\end{equation}
generates  a WIH symmetry itself.

Concluding, {\it if a WIH $(\Delta,[l])$ admits a  non-null vector
field $X$ generating a symmetry, then it admits another symmetry
generator $\phi$ tangent to the leaves of the good cut foliation,
and such that}
\be\label{Xsym} X\ =\ fl +\phi, \ \ \ {\rm where}\ \ \ f={\rm
const}. \ee
{\it The vector field}
\be \hat{\phi}\ =\ \Pi_*\phi \ee
{\it is a  Killing vector field of the metric tensor
$\hat{q}_{AB}$ induced in the space $\hat{\Delta}$ of the null
geodesics in $\Delta$.}

Basically, the argument presented above is the same as that of
\cite{ABL-mech} (due to \cite{LP}), except a small gap in the
proof of (\ref{df}) which was filled in this section.

\section{The phase space of Einstein vacuums admitting  WIH}
As in \cite{ABL-mech}, throughout this paper we will use the covariant
phase space formalism, in a version admitting causal boundaries of
the considered space-time. An exhaustive description of this
formalism can be found for example in \cite{AFK} (see also
\cite{kijowski} and \cite{CJK2} for an alternative framework).

Our covariant phase space ${\bf \Gamma}$ is, briefly speaking, the
space of all the solutions of the vacuum Einstein equations
(without cosmological constant) which admit a non-expanding horizon.
We use the Einstein-Palatini formulation of gravity. The fields
are defined on a given space $M$, a region of a space-time $M'$.
$M$ is contained between three sub-manifolds: $\Delta$,  $M_0$ and
$M_1$, equipped by each point in the phase space with,
respectively,  a non-expanding horizon structure, and space-like
surfaces structures. Below, we first specify the assumptions about
$M$, and secondly the boundary conditions. In Section \ref{Sec:invariance}
we discuss the issue of gauge dependence/invariance of our
boundary conditions and the invariance of the resulting first law
derived in Section \ref{firstlaw}.

$M$ is a closed region of a manifold $M'$, and
\begin{equation}
\dot{M}\ =\ M_0\cup \Delta \cup M_1.
\end{equation}
As in the previous sections, $\Delta=\hat{\Delta}\times[0,1]$
where $\hat{\Delta}$ is a compact, connected $N$ dimensional
manifold. Our calculations will be valid for either of the
following two cases:

\begin{itemize}
\item {\it The properly quasi-local case, Fig. \ref{1}:}
\begin{figure}
\includegraphics[width=8cm]{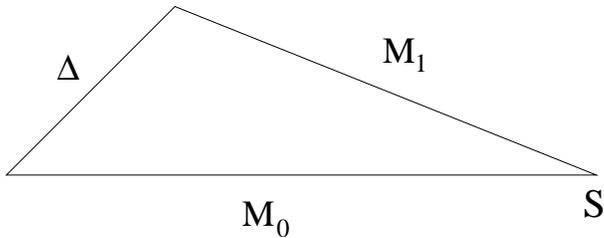}
\caption{\footnotesize{\label{1} The quasi--local case: the space--time
region under consideration is compact and closed,  bounded by
three sub--manifolds with boundary $M_0$, $M_1$ and $\Delta$;
$M_0$ and $M_1$ are bounded by co-dimension 2 compact
sub-manifolds, slices of $\Delta$, and by a single, co-dimension 2,
compact sub-manifold $S$.}}
\end{figure}
The space $M$ is  compact. The surfaces $M_0$ and $M_1$ are
compact, Co-dimension $1$ sub-manifolds with boundary, the boundary
$\partial M_0$ (respectively, $\partial M_1$) consists of an
intersection $\tilde{\Delta}_0$ ($\tilde{\Delta}_1$) with
$\Delta$, and a co-dimension $2$ compact sub-manifold $S$.

\item {\it The asymptotically flat case, Fig. \ref{2}:}
\begin{figure}
\includegraphics[width=10cm]{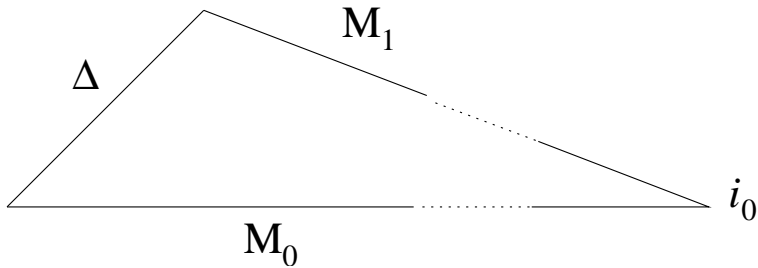}
\caption{\footnotesize{\label{2} The asymptotically flat space--time case:
the space--time region under consideration is infinite, bounded by
sub-manifolds with boundary $M_0$, $M_1$ and $\Delta$; $M_0$ and
$M_2$ are bounded by co-dimension 2 compact sub-manifolds, slices
of $\Delta$; $i_0$ stands for the assumption that the
gravitational fields be asymptotically flat.}}
\end{figure}
$M$ is an infinite region in  $M'$. The surfaces $M_0$ and $M_1$
are co-dimension $1$ infinite  sub-manifolds with boundary. The
boundary $\partial M_0$ (respectively, $\partial M_1$) consists of an
intersection $\tilde{\Delta}_0$ ($\tilde{\Delta}_1$) with
$\Delta$.
 \end{itemize}
In both cases we assume that the metric tensors under
consideration extend smoothly to a neighborhood of ${M}$ in $M'$.

Additionally, in the second case we assume that the metric tensor
fields are asymptotically flat\footnote{ That is, we assume that
the manifold $M$ minus some neighborhood of $\Delta$ is
diffeomorphic to $S_N\times\mathbb{R}\times [0,\,1]$, where $S_N$
is diffeomorphic to an $N$--dimensional sphere, and there exist
spherical coordinates $(x^\ag, r, t)$, flat with respect to some
Minkowski metric tensor $g_{{\rm M}{\mu\nu}}$, such that for large
$r$ the metric tensor components satisfy: $g_{\mu\nu}=g_{{\rm
M}{\mu\nu}} + {\rm O}(\frac{1}{r^{N-1}})$,
 and $\partial_\rho g_{\mu\nu}={\rm
O}(\frac{1}{r^{N}})$. \label{asflat}}.

On the manifold $\Delta$ we fix the Cartesian product structure
$\Delta=\hat{\Delta}\times[0,1]$ and on the manifold
$\hat{\Delta}$ we fix an additional vector field $\hat{\phi}$
later used in the definition of the angular momentum\footnote{ In the 4--dimensional case, when the horizon topology is $S^2$, it was proven
that $\hat\phi$ must generate $SO(2)$ rotations, \emph{i.e.} it must
have closed orbits and its flow must be the identity map for
some value of the flow parameter \cite{ABL-mech}. The normalization
of the vector is later fixed when the resulting angular momentum
formula is compared with Komar's. In our case we cannot exclude 
that our vector field can have more complicated orbits.}.

Out of this data we also construct on $\Delta$ (see Fig. \ref{3}):
\begin{figure}
\includegraphics[width=6cm]{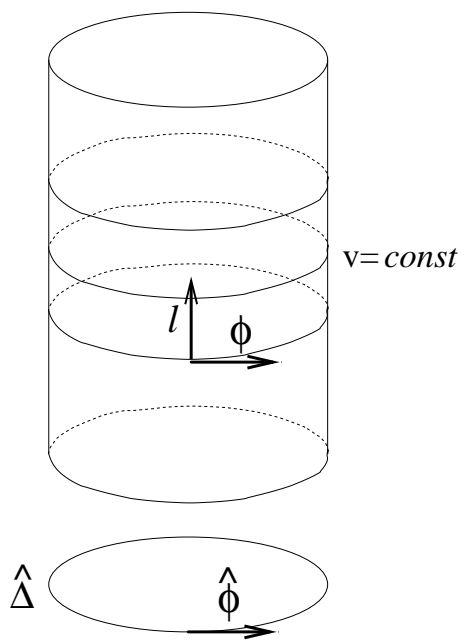}
\caption{\footnotesize{\label{3} The structure fixed in
$\Delta=\hat{\Delta}\times [0,\,1]$: the Cartesian product
structure, the natural coordinate $v$ on $[0,\,1]$ extended to
$\Delta$, the vector field $\frac{d}{dv}$ extended to $\Delta$ and
denoted by $l$, and a vector field $\phi$, the natural extension
to $\Delta$ of a vector field $\hat{\phi}$ fixed on
$\hat{\Delta}$. }}
\end{figure}
$(i)$ a vector field $l$, the natural extension to
$\hat{\Delta}\times[0,1]$ of the vector field $\frac{d}{d v}$,
$v:[0,\,1]\rightarrow [0,\,1]$ being the natural coordinate on the product,
$(ii)$ a function $v:\Delta\rightarrow \mathbb{R}$, the natural
extension of the function $v$ defined on $[0,\,1]$, $(iii)$ a
vector field $\phi$, the natural extension of $\hat{\phi}$ to
$\Delta=\hat{\Delta}\times[0,1]$.

On $M$ we consider vacuum gravitational fields such that $\Delta$
is a non-expanding horizon and $l$ defines a WIH on $\Delta$.
Every gravitational field $g$ is determined by a co-frame,
\emph{i.e.} a sequence of differential 1--forms $(e^1,...,e^{N+2})$
normalized such that
\begin{equation}
g\ =\ -\ e^{N+1}\otimes e^{N+2}\ -e^{N+2}\otimes e^{N+1}\ +\
\hat{\eta}_{AB}\ e^A \otimes e^B,
\end{equation}
where $\hat{\eta}_{AB}$, $A,B=1,...,N$, is the unit matrix 
${\rm diag}(1,...,1)$.
We assume that $M$ is oriented  and  the volume form $e^1\wedge \cdots
\wedge e^{N+2}$ agrees with its orientation. We also assume the
vector field $l$ defined on $\Delta$ to be future pointing.
 We will use
the Palatini framework, therefore we introduce an additional
field, which is an anti-symmetric matrix
$(\Gamma^{\alpha\beta})^{\alpha,\beta=1,...,N+2}$ of 1--forms
referred to as the connection 1--forms.

We formulate now the boundary conditions at $\Delta$ explicitly,
in the technical way. Given a differential n--form $w$ in $M$, its
pullback to $\Delta$ will be denoted by $w_\Delta$.


We assume every co-frame  $(e^1,...,e^{N+2})$ considered in
$M$ and about its dual vector frame $(e_1,...,e_{N+2})$ to satisfy
\ba e_{N+1}\Big| _\Delta\ &=&\ l \label{eq:ebc}\\
\left(e^{N+1}\right)_\Delta\ &=&\ dv \label{eq:ebc1}\\
\left(e^{N+2}\right)_\Delta\ &=&\ 0 \label{eq:ebc2}. \ea

The pullbacks to $\Delta$ of  the connection 1--forms are subject
to the following conditions:
\ba \left(\Gamma^{N+2}{}_1\right)_{\Delta} = \ldots =
\left(\Gamma^{N+2}{}_N\right)_{\Delta}\ &=&\ 0,\label{eq:gbc2}\\
{\cal L}_l \left(\Gamma^{N+1}{}_{N+2}\right)_{\Delta}\ &=&\
0\label{eq:gbc1}. \ea
%
The conditions (\ref{eq:ebc}, \ref{eq:ebc2}) imply that $\Delta$ is
null, the condition (\ref{eq:gbc2}) is equivalent to the
assumption that $\Delta$ be a non-expanding horizon contained in a
vacuum space-time (it is exactly equivalent to (\ref{Llq})),
whereas (\ref{eq:gbc1}) (meaning the same as (\ref{Llomega})) is 
the necessary and
sufficient condition for the vector field $l$ to form 
a WIH together with $\Delta$.

Just to simplify the calculations we introduce a short--hand
notation for the following $N+2-k$--forms
\be
 \Sigma_{\underbrace{\alpha\beta...\gamma}_{k}} = \frac{1}{(N-k+2)!}
\epsilon_{\ag\bg...\cg\dg...\eg}\,\underbrace{e^\dg\wedge \cdots \wedge
e^\eg}_{N+2-k}, \label{eq:DefSigma} \ee
and we denote the curvature of $\Gamma^{\alpha}\!_{\beta}$
\be F^{\ag\bg}\ :=\ d\Gamma^{\ag\bg} +\Gamma^{\ag}\!_{\cg}\wedge\Gamma^{\cg\bg}
\label{eq:DefF}.\ee
The Palatini action for the vacuum Einstein equations in arbitrary
dimension can be written down as
\be S = C \int_M F^{\ag\bg}\wedge \Sigma_{\ag\bg}\ +\ S_\partial
\ee
with $C$ being a constant of dimension $L^{N}$.
The boundary term $S_\partial$ is unnecessary in the
properly quasi--local case
\begin{equation}
S_\partial\ =\ 0
\end{equation}
whereas in the asymptotically flat case we take it to be
\begin{equation}
S_\partial\ =\ -C\lim_{r\rightarrow\infty}\int_{\tau_r}
 \Gamma^{\ag\bg}\wedge \Sigma_{\ag\bg}
\end{equation}
where $\tau_r$ stands for the cylinder (a sphere world-sheet)
$r={\rm const}$ (see Footnote \ref{asflat}). That additional
boundary term is added to ensure that the variational problem with
asymptotically flat boundary conditions is equivalent to the
vacuum Einstein equations. Its role can be seen easily if we
compute the variation of $S$ corresponding to an arbitrary vector
field $\delta$ tangent to the phase space ${\bf \Gamma}$, namely
\be \delta S =\int_M S_{\Sigma_{\mu\nu}}\delta \Sigma_{\mu\nu} +
\int_M S_{\Gamma_{\mu\nu}}\delta \Gamma_{\mu\nu} - C
\lim_{r\rightarrow\infty}\int_{\tau_r} \Gamma^{\ag\bg}\wedge \delta
\Sigma_{\ag\bg} \ee
Now, it follows from the asymptotic flatness that the boundary
term is zero, whereas the vanishing of the bulk terms is
equivalent to the Einstein equations.

It is easy to check that in neither case we need to add any
surface term associated with $\Delta$. This is a consequence of the
properties of non-expanding horizons \cite{AFK}, \cite{ABL-mech}.

The covariant phase space is equipped with a pre-symplectic
structure, \textit{i.e.} an antisymmetric 2--form ${{\bf
\Omega}}$. It is defined by a pre-symplectic current calculated in
the usual way: take two vector fields $\delta_1$ and $\delta_2$
tangent to ${\bf \Gamma}$, and consider the following identity
\be \delta_1\,\delta_2\,S - \delta_2\,\delta_1\,S -
\left[\delta_1,\delta_2\right] S = 0. \ee
If the Einstein equations are satisfied, the left hand side takes
the following form
\be \delta_1\,\delta_2\,S - \delta_2\,\delta_1\,S -
\left[\delta_1,\delta_2\right] S\ =\ -C\int_{\partial M}\delta_1
\Gamma^{\ag\bg}\wedge \delta_2 \Sigma_{\ag\bg}\ -\ \delta_2
\Gamma^{\ag\bg} \wedge \delta_1 \Sigma_{\ag\bg}. \label{eq:ident}\ee
The boundary of $M$ consists of the two space--like hyper-surfaces
$M_{0}$ and $M_{1}$, the horizon $\Delta$,  plus, in the
asymptotically flat case, the cylinder $\tau_r$ where
$r\rightarrow  \infty$.

The differential $(N+1) $--form equal to $C$ times the integrant
is called the pre--symplectic current
\begin{equation}
j(\delta_1,\delta_2)\ =\ -C \left(\delta_1 \Gamma^{\ag\bg}\wedge
\delta_2 \Sigma_{\ag\bg}\ -\ \delta_2 \Gamma^{\ag\bg} \wedge
\delta_1 \Sigma_{\ag\bg}\right).
\end{equation}

\begin{figure}
\includegraphics[width=10cm]{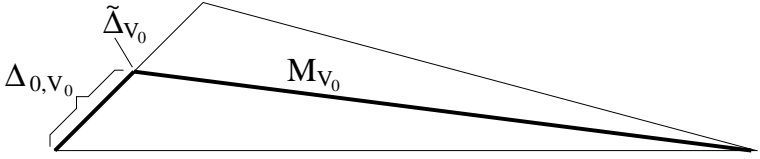}
\caption{\footnotesize{\label{4} The integration surface for the
symplectic form}}
\end{figure}

The pre-symplectic form may be defined by integrating the
pre-symplectic current
\begin{equation}\label{Omega}
{\bf \Omega}(\delta_1,\delta_2)\ =\ \int_{\Delta_{0,v_0}\cup
M_{v_0}}j(\delta_1,\delta_2)
\end{equation}
along any of the surfaces ${\Delta_{0,v_0}\cup M_{v_0}}$ (see Fig.
\ref{4}) where $\Delta_{0,v_0}$ is the portion of the surface
$\Delta$ bounded by the slices $\tilde{\Delta}_0$ such that $v=0$,
 and $\tilde{\Delta}_{v_0}$ such that  $v=v_0$,
 \begin{equation}
 \Delta_{0,v_0}\ :=\ \hat{\Delta}\times [0,\,v_0].
 \end{equation}
$M_{v_0}$ is an $N+1$ dimensional sub-manifold of $M'$ whose
boundary consists of $\tilde{\Delta}_{v_0}$ and: either $S$ in
the quasi-local case or the `sphere at infinity' in the
asymptotically flat case with $M_{v_0}$ being asymptotically flat
in the asymptotically flat coordinates (see Footnote
\ref{asflat}).  Due to the identity (\ref{eq:ident}), the 
integral (\ref{Omega}) is
independent of the choice of surface labelled by any $0\le v_0\le
1$. Also the domains of dependence of each of the surfaces which
can be used in (\ref{Omega}) are equal to the domain of dependence
of $M_0$. In particular, we could choose $v_0=0$ and just $M_0$ as
the integration surface. Therefore, the integral   depends only on
the definition of region $M$ and, in particular, on the choice
of surface $M_0$ itself. Note that by considering an arbitrary
value of $v_0$ in (\ref{Omega}) we will be able to see explicitly
what that dependence is like, and why our final results are
independent of that choice.

In the integral defining the pre-symplectic form
(\ref{Omega}) the orien\-tation of the surface ${\Delta_{0,v_0}\cup
M_{v_0}}$ is the one given by the Stokes' theorem applied to  the
region bounded by $M_0$ and ${\Delta_{0,v_0}\cup M_{v_0}}$. In
particular,  the volume forms on $\Delta_{0,v_0}$ and,
respectively, $M_{v_0}$ are: $-dv\wedge e^1\wedge\cdots\wedge
e^N$, and $e^{N+2}\wedge e^1 \wedge \cdots\wedge e^N$.

We will analyze the horizon part of the symplectic form first.

The boundary conditions (\ref{eq:ebc}, \ref{eq:ebc1},
\ref{eq:ebc2}, \ref{eq:gbc2}, \ref{eq:gbc1}) imply that, in particular,
 given a gravitation field, the rotation 1--form potential
$\omega$ of the WIH $(\Delta,[l])$ defined by the vector field $l$
is given by the following pullback to he horizon
\begin{equation} \left(\Gamma^{N+1\,N+2}{}\right)_\Delta = -\omega.
\end{equation}
The induced null surface geometry of $\Delta$ determines the area
$N$--form $\eps$ which can be expressed as
\be \left(\Sigma_{N+1\,N+2}{}\right)_\Delta =
\left(e^1\wedge\cdots\wedge e^N\right)_\Delta\ =:\ -\eps. \ee The
area of arbitrary space-like $N$-dimensional hyper-surface
$\tilde{S}\subset\Delta$ is equal to $\int_{\tilde{S}}\eps$
(provided that the orientation of $\tilde{S}$ is fixed appropriately).
On the other hand,
 \ba
 \Sigma_{N+1\,A} &=& \pm \underbrace{ e^1\wedge \cdots \wedge e^N }_\textit{no A}
 \wedge e^{N+2} \\
 \Sigma_{N+2\,A} &=& \pm \underbrace{ e^1\wedge \cdots \wedge e^N}_\textit{no A}
 \wedge e^{N+1} \\
 \Sigma_{AB} &=& \pm \underbrace{ e^1 \wedge \cdots \wedge e^N }_\textit{no A and B}
 \wedge e^{N+1} \wedge e^{N+2}
\ea
 Note that since $\Sigma_{N+1\,A}$ and $\Sigma_{AB}$ contain $e^{N+2}$,
 they must vanish when pulled back to $\Delta$. We are left therefore
 with
\be \int_{\Delta_{0,v_0}} j(\delta_1,\delta_2) = - 2C
\int_{\Delta_{0,v_0}}\,\delta_1 \omega \wedge \delta_2 \eps -
 \delta_2 \omega \wedge \delta_1 \eps
\ee

We can simplify this expression if we decompose $\omega$ to \be
\omega = \kappa^{(l)} dv + \tilde \omega \ee with $l \lrcorner
\tilde\omega = 0$. It is straightforward to prove that
$\tilde\omega\wedge\eps = 0$ and therefore the expression for the
horizon part of the symplectic form can be further simplified
\begin{equation} \int_{\Delta_{0,v_0}} j(\delta_1,\delta_2) =  -2C \left(
 \delta_1\kappa^{(l)} \int_{\Delta_{0,v_0}}\,
 \,d\left(v \delta_2\eps\right) - \delta_2\kappa^{(l)}
 \int_{\Delta_{0,v_0}}\,
 \,d\left(v \delta_1\eps\right)\right)
\end{equation}
The horizon part of the identity can finally be integrated out to
\be
 \int_{\Delta_{0,v_0}} j(\delta_1,\delta_2) =
 -2v_0 C\left( \delta_1 \kappa^{(l)} \delta_2 a_\Delta
 \ -\ \delta_2 \kappa^{(l)} \delta_1 a_\Delta \right)
\ee
where $a_\Delta$ stands for the area of a cross-section of
$\Delta\rightarrow \hat{\Delta}$.

We conclude that the symplectic form reads
\ba {\bf \Omega}(\delta_1,\delta_2) &=& - 2Cv_0\,
\left(\delta_1\kappa^{(l)}
 \delta_2 a_\Delta - \delta_2\kappa^{(l)}
 \delta_1 a_\Delta\right)\ +\nonumber\\ &{}&-\ C\int_{M_{v_0}}
\left(\delta_1\Gamma^{\ag\bg}\wedge\delta_2\Sigma_{\ag\bg} -
\delta_2\Gamma^{\ag\bg}\wedge\delta_1\Sigma_{\ag\bg}\right) \nonumber\\
 &:=&\ {\bf \Omega}_{\Delta_{0,v_0}}(\delta_1,\delta_2)\ +\
 {\bf \Omega}_{\rm bulk}(\delta_1,\delta_2),
\ea
where ${\bf \Omega}_{\rm bulk}$ (the `bulk' part) is the integral
and the remaining (first) term constitutes ${\bf
\Omega}_{\Delta_{0,v_0}}$ (the `horizon' part). The horizon part
is proportional to $v_0$, the distance in terms of the affine
parameter corresponding to $l$, between the slices
$\tilde{\Delta}_{v_0}$ and $\tilde{\Delta}_0$. The horizon part is
invariant with respect to rescalings of $l$ by a constant, but it
depends on the choice of the initial slice $\tilde{\Delta}_0$.

\section{Generating functions for space--time diffeomorphisms}
A special class of vector fields tangent to the phase space ${\bf
\Gamma}$ is defined by the diffeomorphisms of $M$. Assign
a vector field $X$ defined in a neighbourhood of $M$ to every point in
${\bf \Gamma}$ (a solution of the Einstein equations on $M$). From our point
of view, the flow of $X$ transports the gravitational field, while
the region $M$ is kept fixed. A necessary condition for $X$ is
that the flow of $X$ understood in this way preserves the boundary
conditions (\ref{eq:ebc}, \ref{eq:ebc1}, \ref{eq:ebc2},
 \ref{eq:gbc2}, \ref{eq:gbc1}) at $\Delta$.

 The Lie derivative along $X$ defines a
 vector field $\delta_X$ tangent to ${\bf \Gamma}$ (or a variation, as this object is
often called in variational calculus). Our goal is to formulate
conditions upon which the flow of $\delta_X$ is generated by a
 Hamiltonian function. Namely, we want to determine the conditions under which
 there exists a function $H$ on ${\bf \Gamma}$ such that
 \be
  -\delta H = {\bf {\bf \Omega}}(\delta_X,\delta) \label{eq:deltaH}
 \ee
 for every vector field $\delta$ (variation) tangent to ${\bf \Gamma}$
 (in other words $-dH=\delta_X\lrcorner{\bf \Omega}$, where all the operations
 apply to differential forms and vectors in ${\bf \Gamma}$.)

 Before we proceed,  we introduce several formulas we will use in subsequent
 calculations.

We assume the vacuum Einstein equations to be satisfied on entire
space-time
 \ba
  D\Sigma_{\ag\bg} &\equiv& d\Sigma_{\ag\bg} - \Gamma^\cg\!_\ag \wedge \Sigma_{\cg\bg}
- \Gamma^\cg\!_\bg \wedge \Sigma_{\ag\cg} =  0 \label{eq:DSigma}\\
  F^{\ag\bg} \wedge \Sigma_{\ag\bg\cg} &=& 0.
 \ea
 Those imply following identities
 for the variations
 \ba
 D\,\delta\Sigma_{\ag\bg} &=& \delta \Gamma^\cg\!_\ag \wedge \Sigma_{\cg\bg} -
  \delta \Gamma^\dg\!_\bg \wedge \Sigma_{\dg\ag} \\
 \delta F^{\ag\bg} \wedge \Sigma_{\ag\bg\cg} &=& - F^{\ag\bg}\wedge
\delta e^\dg
 \wedge \Sigma_{\ag\bg\cg\dg} \label{eq:deltaF}.
 \ea
 These are in fact the linearized Einstein equations.

 Two useful identities follow from (\ref{eq:DefSigma}):
\be
 \delta \Sigma_{\ag\bg\cg} = \delta e^\dg \wedge\Sigma_{\ag\bg\cg\dg}
\ee
and
\be
 X \lrcorner \Sigma_{\ag\bg} = X^\cg\,\Sigma_{\ag\bg\cg} \label{eq:Xwedge}
\ee
with $X^\cg = e^\cg (X)$.

 We now calculate ${\bf \Omega}_{\rm bulk}(\delta_X,\delta)$.
 The variations of the fields are equal to the Lie
 derivative along $X$. By applying the well--known Cartan formula
${\cal L}_X\alpha = d(X\lrcorner\alpha) + X\lrcorner d\alpha$
 we get
 \ba
 {\bf \Omega}_{\rm bulk}(\delta_X,\delta) = -C
 \int_{M_{v_0}} &&\left( d(X\lrcorner\Gamma^{\ag\bg})\wedge
 \delta\Sigma_{\ag\bg} + (X\lrcorner d\Gamma^{\ag\bg})\wedge \delta
 \Sigma_{\ag\bg}\right. - \\
&&\left. -  \delta\Gamma^{\ag\bg}\wedge d(X\lrcorner\Sigma_{\ag\bg})
 - \delta\Gamma^{\ag\bg}\wedge (X \lrcorner d\Sigma_{\ag\bg})\right).
\label{eq:OmegaX}
\ea
 This formula can be reduced to mere boundary terms.
 First we apply integration by parts to the first and third term,
 the definition of curvature 2--form (\ref{eq:DefF}) to the second
 and (\ref{eq:DSigma}) to the fourth. We get
\ba
 {\bf \Omega}_{\rm bulk}(\delta_X,\delta)\ &=&\ -C
 \int_{\partial M_{v_0}} \left( (X\lrcorner\Gamma^{\ag\bg})\delta
\Sigma_{\ag\bg} + \delta\Gamma^{\ag\bg}\wedge(X\lrcorner\Sigma_{\ag\bg})\right)
+\\
&&-C\int_{M_{v_0}} \left(-\delta
F^{\ag\bg}\wedge(X\lrcorner\Sigma_{\ag\bg}) +(X \lrcorner
F^{\ag\bg})\wedge\delta\Sigma_{\ag\bg}\right). \ea

The remaining bulk term can be proved to vanish. By the virtue of
(\ref{eq:Xwedge}) and (\ref{eq:deltaF}) we have
\be
-\delta F^{\ag\bg}\wedge(X\lrcorner\Sigma_{\ag\bg}) = - X^\cg \delta F^{\ag\bg}
\wedge\Sigma_{\ag\bg\cg} = X^\cg F^{\ag\bg} \wedge \delta e^\dg\wedge
\Sigma_{\ag\bg\cg\dg}.
\ee
The bulk term can be now rewritten in the form of
\be
-\delta F^{\ag\bg} \wedge(X\lrcorner\Sigma_{\ag\bg}) + (X\lrcorner F^{\ag\bg})
\wedge\delta\Sigma_{\ag\bg} = \delta e^\dg\wedge (X^\cg F^{\ag\bg}\wedge
\Sigma_{\ag\bg\cg\dg} - (X\lrcorner F^{\ag\bg})\wedge\Sigma_{\ag\bg\dg}).
\ee
The last term is just the contraction of the
right--hand side of Einstein equations (\ref{eq:deltaF})
with $X$ and is therefore equal to 0.\footnote
{
All calculations shown here do not differ significantly from
  the four--dimensional case. This may be traced back to the fact that
  in all wedge products of forms the first forms are either
  the curvature 2--form $F^{\ag\bg}$ or the connection 1--form
  $\Gamma^{\ag\bg}$.
} We are left only with the surface terms of the bulk part of the
symplectic form
\be {\bf \Omega}_{\rm bulk}(\delta_X,\delta)\ =\ -C \int_{\partial
M_{v_0}} \left( (X\lrcorner \Gamma^{\ag\bg}) \delta\Sigma_{\ag\bg}
+ \delta\Gamma^{\ag\bg} \wedge(X\lrcorner \Sigma_{\ag\bg})\right)
\ee.
The horizon part of the symplectic form, on the other hand,
vanishes,
\be {\bf \Omega}_{\Delta_{0,v_0}}(\delta_X,\delta)\ =\ -2Cv_0
\left( \Lie_X \kappa^{(l)}\delta a_\Delta -
\delta\kappa^{(l)}\Lie_X a_\Delta\right) \ =\ 0, \ee because both
the area $a_\Delta$ and the surface gravity of $l$ are constant on
a WIH.

Finally
\be\label{eq:sympformhor} {\bf \Omega}(\delta_X,\delta)\ =\ -C
\int_{\partial M_{v_0}} \left( (X\lrcorner \Gamma^{\ag\bg})
\delta\Sigma_{\ag\bg} + \delta\Gamma^{\ag\bg} \wedge(X\lrcorner
\Sigma_{\ag\bg})\right). \ee
Every vector field  $X$ can be split into two parts, say $X'$ and
$X''$, one being identically zero outside some finite neighborhood
of the horizon while the other identically vanishing at the
horizon. In the asymptotically flat case, the contribution of
$X''$ to (\ref{eq:sympformhor}) is the variation of some ADM
momentum (see \cite{ABL-mech}), provided that $X''$ satisfies
appropriate conditions in the infinity. $X'$, on the other hand,
is responsible for the horizon contribution to (\ref{eq:sympformhor}).

Our goal is to derive the contribution from the horizon. We assume
$X$ is identically zero out of a finite neighborhood of the
horizon, and in the properly quasi-local case we assume the
support ${\rm supp}(X)$ intersects the boundary of $M_{v_0}$ at
$\Delta$ only,
\be {\rm supp}(X)\cap \delta M_{v_0}\subset\Delta.
\ee
Suppose that $X$ satisfies the following boundary
condition on the horizon:
\be X\Big|_\Delta\ =\ \frac{\kappa^{(X)}}{\kappa^{(l)}} \,l -
\Omega^{(X)} \,\phi, \label{Xih}\ee
where the quantities $\kappa^{(X)}$ and $\Omega^{(X)}$ are
constant on the horizon, but their values possibly depend on the
gravitational field\footnote{$X$ is well defined for the
gravitational fields such that $\kappa^{(l)}\not=0$. In the $4$
(space--time) dimensional case the resulting Hamiltonian may be extended
by continuity to the extremal points of ${\bf \Gamma}$ as well.}.
That is $\kappa^{(X)}$ and $\Omega^{(X)}$ are functions defined on
the phase space ${\bf \Gamma}$. (One should not confuse
$\Omega^{(X)}$ with the symplectic form.) The reason for this
notation as well as the geometric and physical meaning of the two
quantities introduced here will be explained in the next three
sections. The form (\ref{Xih}) of $X$ at $\Delta$ is a natural
assumption in the case of a  WIH admitting a non-null symmetry
generator (compare with (\ref{Xsym})). We discuss this case in
Section \ref{Sec:invariance}.

By substituting $X$ in (\ref{eq:sympformhor}) by the right hand
side of (\ref{Xih})  and taking into account the boundary
conditions at $\Delta$ ((\ref{eq:ebc}), (\ref{eq:ebc1}) and
(\ref{eq:ebc2})) we get\footnote{We use below the following
identity satisfied by the pullbacks of the involved differential
forms onto $\tilde{\Delta}_{v_0}$: $\left(\omega\wedge
\epsilon\right)_{\tilde{\Delta}_{v_0}} = 0$.}
\ba {\bf \Omega}(\delta_X,\delta)\ &=&\
-2C\int_{\tilde{\Delta}_{v_0}}\left((X\lrcorner\omega)\delta\eps
+\delta\omega\wedge(X\lrcorner \eps)\right)\nonumber\\
&=&\ -2C\int_{\tilde\Delta_{v_0}}\left(\kappa^{(X)}\delta\eps
-\Omega^{(X)}(\phi\lrcorner\omega)\delta\eps -
\Omega^{(X)}(\phi\lrcorner\delta\omega)\eps \right)\nonumber\\ &=&\
-2C\int_{\tilde\Delta_{v_0}} \left( \kappa^{(X)}\,\delta \eps
-\Omega^{(X)}\delta\left((\phi \lrcorner \omega)\eps\right)
\right)\nonumber\\
&=&\ 2C\,\kappa^{(X)}\delta a_\Delta + 2C \,\Omega^{(X)}
\delta\left(\int_{\tilde{\Delta}_{v_0}} (\phi \lrcorner
\omega)\eps\right)\label{eq:AA} \ea
The  sign of the first term of the result follows from the
orientation of $\tilde{\Delta}_{v_0}$ defined by the Stokes'
theorem, in which $\epsilon$ is minus the
$N$-volume element.
\section{Angular momentum}\label{angmom}

We now assume that the vector field $X$ coincides with
the fixed vector field  $\phi$ on the horizon, \emph{i.e.} 
$\Omega^{(X)} = -1$ and
$\kappa^{(X)} = 0$. Then, ${\bf \Omega}(\delta_X,\delta)$
calculated in (\ref{eq:AA}) is necessarily a variation\be
{\bf \Omega}(\delta_\phi,\delta) = -2C \int_{\tilde\Delta_{v_0}}
\delta(\phi\lrcorner \omega) \eps =: -\delta J^{(\phi)}, \ee
where the generator
\be\label{J} J^{(\phi)}\ :=\ 2C\int_{\tilde\Delta_{v_0}} (\phi
\lrcorner \omega) \eps \ee
will be called the WIH \emph{angular momentum}  associated to the
vector field $\phi$ tangent to $\Delta$. (We recall that the
orientation of $\tilde{\Delta}_{v_0}$ is such that $\epsilon$ is
minus the area form.) Due to the fact that $\phi\lrcorner\omega$
is a function constant along the null geodesic generators of
$\Delta$ (for both $\phi$ and $\omega$ are Lie dragged by $l$),
and the area $N$--form is both Lie dragged by and orthogonal to
$l$, the horizon angular momentum $J^{(\phi)}$ is independent of
$v_0$. The formula agrees with the one given in \cite{ABL-mech} in
the case of four--dimensional horizons.

\section{Area as a generator of null translations}
If we take $X$ to be just a null vector
proportional to $l$ at $(\Delta, [l])$, \emph{i.e.} 
$\Omega^{(X)} = 0$, $\kappa^{(X)}
\neq 0$, we get from (\ref{eq:AA})
\be -\delta H_{X} = 2C\kappa^{(X)} \delta a_{\Delta}. \ee
The necessary condition for the flow to be Hamiltonian 
is that $H_X$ and $\kappa^{(X)}$ be
functions of $a_\Delta$ solely. Conversely, given any function $H$
of the area, one can construct the corresponding Hamiltonian vector
field $\delta_X$ in the following way. Fix a vector field $Y$ in
$M$, such that $Y\Big|_\Delta\ =\ l$, and $Y$ vanishes out of a
sufficiently small neighborhood of $\Delta$. Given a gravitational
field, \emph{i.e.} a point in ${\bf \Gamma}$, define in $M$ a vector field
$X$,
\be X\ :=\  -\frac{1}{2C\kappa^{(l)}}H'\,Y,\ee
where $H'$ is the derivative of $H$. Then, the corresponding
vector field $\delta_X$ tangent to ${\bf \Gamma}$ satisfies
\be -\delta H\ =\ {\bf \Omega}(\delta_X,\delta)\ee
for every vector field $\delta$ tangent to ${\bf \Gamma}$.

In particular, the horizon area itself is a generator of a
semigroup of diffeomorphisms, whose flow of the horizon $\Delta$ is
defined  by the null vector field $(-2C\kappa^{(l)})^{-1} l$. In
other words, the horizon area is a generator of the null
translations of the horizon, as it was first observed in the case
of the Killing horizons in $4$-space-time dimensions in
\cite{Wald} and shown for WIHs in $4$ space-time dimensions in
\cite{ABL-mech}.

\section{The first law}
\label{firstlaw} In this section we will show that the requirement
that the evolution of the gravitational field be Hamiltonian for
general vector fields $X$ satisfying (\ref{Xih}) implies a first
law of black hole thermodynamics analogous to the case of four
space-time dimensions.

Assume the left hand side of (\ref{eq:AA}) \emph{is} a total
variation of a quantity $H_{\Delta}$. The equality reads
\be\label{First} -\delta (H_{\Delta}) = 2C \kappa^{(X)}\,\delta
a_{\Delta} + \Omega^{(X)}\,\delta J^{(\phi)}. \ee
Necessary conditions are:
\begin{itemize}
\item $H_{\Delta}$, $\kappa^{(X)}$ and
$\Omega^{(X)}$ are arbitrary functions of $a_{\Delta}$ and
$J^{(\phi)}$ only,
\item
 an integrability condition analogous to the Maxwell relation in
 thermodynamics holds:
\be
 2C\frac{\partial \kappa^{(X)}}{\partial J^{(\phi)}} =
\frac{\partial \Omega^{(X)}}{\partial a_{\Delta}}. \ee
\end{itemize}
Equation (\ref{eq:deltaH}) imposes therefore strong constraints on
the coefficients of the vector field $X$ at $\Delta$.

The dependence of $H_\Delta$ on $a_\Delta$ and $J^{(\phi)}$ is
arbitrary. In the $4$-space-time dimensional case some additional
conditions on the vector field $X$ were  imposed \cite{ABL-mech},
which eventually determined the generating function $H_\Delta$
completely. Firstly, it was not assumed that $X$ was zero except for a
neighborhood of $\Delta$. Instead, it was  assumed that at the
spacial infinity $X$ was a normalized generator of a time
translation. Then, the corresponding function $H_X$ consists of
two contributions, one from $\Delta$ and another from infinity, namely
\be H_X = H_{\rm ADM}\ -\ H_\Delta, \ee
where $H_{\rm ADM}$ is the ADM mass. Secondly,  it was assumed
that $X$ was assigned to each gravitational field, point in our
phase space ${\bf \Gamma}$, in such a way that, whenever the
space-time was asymptotically flat {\it and stationary}, $X$
coincided with the Killing vector field. If this assumption can be
satisfied, then
\be\delta_X\ =\ 0, \ \ \ \ \ \ \ \ {\bf \Omega}(\delta_X,\delta)\
=\ 0,\ee
and, in  consequence
\be \delta(H_\Delta)\ =\ \delta(H_{\rm ADM}) \ee
for every stationary gravitational field. The two equalities
determine $H_\Delta$ by the ADM mass of the Kerr solution, modulo
a constant. This is possible due to the uniqueness of a
stationary, asymptotically flat, vacuum black hole of a given area
$a_\Delta$ and angular momentum $J^{(\phi)}$. Since there is no
natural constant of the right units in the vacuum case, the
undetermined constant  was set to zero. Remarkably, the resulting
$H_\Delta$ extends smoothly to the extremal WIH case. (Incidently,
it was not proven that a suitable assignment of $X$ to every
point ${\bf \Gamma}$ really exits. The conditions used were
necessary only.)

As we mentioned in the introduction, the uniqueness does not any
longer hold for asymptotically flat, axially symmetric vacuum
space-times  in the general, $N+2$-dimensional case. The issue is
similar to that of the hairy black holes \cite{acs} in $4$
dimensions.
 
However, if one restricts the phase space ${\bf \Gamma}$ to
spherically symmetric (static) gravitational fields
\cite{Rogatko1}, \cite{Rogatko2}, \cite{Rogatko3},
then, just like in the four--dimensional case,  the value of energy $H_\Delta$
as a function of horizon area can be determined. It is given by
the function describing the ADM mass as a function of the horizon
area (see \cite{MyersPerry})
%
%
\be
H_\Delta = CN\left(\frac{2\pi^{\frac{N+1}{2}}}{\Gamma\left(\frac{N+1}{2}\right)}\right)^{\frac{1}{N}} A^{1-\frac{1}{N}}
\ee
with $\Gamma(x)$ denoting Euler's gamma function. 

\section{Invariance and non-invariance}\label{Sec:invariance}
Let us discuss now the gauge invariance of our framework. In order to
define the angular momentum and derive first law (\ref{First}) we
have equipped a non-expanding horizon $\Delta$ with extra
structure: a WIH flow $[l]$, a foliation, a representative
$l\in[l]$ and a vector field $\hat{\phi}$ defined on the space
$\Delta$ of the null generators of $\Delta$. Also, the transversal
surface $M_0$ was used  for the definition of the pre-symplectic
form. It is easy to observe that the generator of the null
translations of $\Delta$, the area functional $a_\Delta$ is in
fact uniquely defined given a non-expanding horizon only, and it
is determined just by any cross-section. The definition of the
second relevant parameter, the angular momentum $J^{(\phi)}$,
involves all the elements of extra structure. We show below,
however, that if the vector field $\hat{\phi}$ generates a local
symmetry of the area element induced in $\hat{\Delta}$, then a
corresponding $J^{(\phi)}$ depends only on the geometry  of
$\Delta$ and on the vector field $\hat{\phi}$ itself. And again it
is determined by an arbitrary cross-section of $\Delta$. Indeed,
the integral (\ref{J}) can be written as an integral along the
base manifold $\hat{\Delta}$, namely
\be J^{(\phi)}\ =\ -2C\int_{\hat{\Delta}} (\hat{\phi} \lrcorner
\hat{\omega}) \hat{\eps} \ee
where $\hat{\eps}$ is the area element defined on $\hat{\Delta}$
by the induced metric tensor $\hat{q}$ and  $\hat{\omega}$ is the
pullback of the rotation 1--form potential $\omega$ by the map
\be \hat{\Delta}\rightarrow \tilde{\Delta}_{v_0} \ee
naturally defined, given a cross-section $\tilde{\Delta}_{v_0}$.
Now, given a non-expanding horizon  geometry on $\Delta$, a change
of the WIH flow $[l]$ and the foliation, both used to define
$\hat{\omega}$, amounts to a transformation
\be \hat{\omega}\ \mapsto\ \hat{\omega} + dg,\ \ \ \ \ \ {\rm
where}\ \ \ \ \ \ \ g:\hat{\Delta}\rightarrow \mathbb{R}.\ee
Consequently
\ba \int_{\hat{\Delta}}(\hat{\phi}\lrcorner\hat{\omega}')
\hat{\eps}\ &=&\
 \int_{\hat{\Delta}}\hat{\omega}'\wedge (\hat{\phi}\lrcorner\hat{\eps})\nonumber\\
\ &=&\ \int_{\hat{\Delta}}(\hat{\omega}+dg)\wedge (\hat{\phi}\lrcorner\hat{\eps})\nonumber\\
&=&\ \int_{\hat{\Delta}}\left(\hat{\omega}\wedge
(\hat{\phi}\lrcorner\hat{\eps})\ +\ g
(\Lie_{\hat{\phi}}\hat{\eps})\right)\nonumber
\\
&=& \int_{\hat{\Delta}}(\hat{\phi}\lrcorner\hat{\omega})
\hat{\eps} \ +\ \int_{\hat{\Delta}} g \Lie_{\hat{\phi}}\hat{\eps}.
\ea
Therefore, in the most elegant formulation of this framework one can
add one more boundary condition, namely the assumption that for
every gravitational field in ${\bf \Gamma}$ the WIH defined on
the surface $\Delta$ by the vector field $l$ admits a non-null
symmetry generator $Y$ such that $\Pi_*Y$ is a fixed vector field
$\hat{\phi}$ on $\hat{\Delta}$. A generic WIH of this type admits
exactly a 2-dimensional group of symmetry generators
\ref{Sec:sym}. Then, our assumption (\ref{Xih}) is equivalent to
assuming that the vector field $X$ restricted to $\Delta$ is a
symmetry generator.



\section{Discussion and conclusions}
\label{discussion} We proved that the first law of black hole
thermodynamics holds for arbitrarily dimensional
weakly isolated horizons. We have defined the horizon angular
momentum as the generator of rotational symmetry. We have shown
that the assumption that time flow on the horizon is Hamiltonian
leads to a differential condition on the mass function which can
be interpreted as the first law of thermodynamics of black holes.

We also proposed a Hamiltonian function depending on the horizon area and
angular momentum.

\section{Acknowledgments}

We would like to thank Abhay Ashtekar and Jerzy Kijowski for
discussions. This work was supported in part by the Polish
Committee for Scientific Research (KBN) under grants no: \mbox{2
P03B 127 24}, \mbox{2 P03B 130 24}, the National Science
Foundation under grant  0090091, and the Albert Einstein Institute
of the Max Planck Society.


\begin{thebibliography}{99}



\bibitem{LP} J. Lewandowski, T. Paw\l{}owski, Geometry of non--expanding
horizons: n--dimensional generalization, gr-qc/0410146 (2004)

\bibitem{AFK}
A. Ashtekar, S. Fairhurst, B. Krishnan,  Isolated Horizons:
Hamiltonian Evolution and the First Law, {\em Phys. Rev. }
{\bf D62} 104025 (2000), gr-qc/0005083

\bibitem{Aetal} A. Ashtekar, C. Beetle, O. Dreyer, S. Fairhurst, B. Krishnan,
J. Lewandowski, J. Wi\'sniewski, Generic Isolated Horizons and Their
Applications, \emph{Phys. Rev. Lett.} {\bf 85} 3564-3567 (2000),
 gr-qc/0006006

\bibitem{ABL-ge} A. Ashtekar, C. Beetle, J. Lewandowski,
 Geometry  of generic isolated horizons, \emph{Class. Quantum
Grav.} {\bf 19} 1195-1225 (2002), gr-qc/0111067

\bibitem{ABL-mech}
A. Ashtekar, C. Beetle, J. Lewandowski,  Mechanics of Rotating Isolated
Horizons, \emph{Phys. Rev.} {\bf D64} 044016 (2001), gr-qc/0103026

\bibitem{Rogatko1}
M. Rogatko,  Uniqueness Theorem for Static Black Hole Solution of
 $\sigma$--models in Higher Dimensions, \emph{Class. Quantum Grav.}
{\bf 19} L151 (2002), hep-th/0207187

\bibitem{Rogatko2}
M. Rogatko, Uniqueness Theorem of Static Degenerate and Non--degenerate
Charged Black Holes in Higher Dimensions, \emph{Phys. Rev.}
{\bf D67} 084025 (2003), gr-qc/0302091

\bibitem{Rogatko3}
M. Rogatko,  Uniqueness Theorem for Generalized Maxwell Electric and
Magnetic Black Holes in Higher Dimensions, \emph{Phys. Rev.}
{\bf D70} 044023 (2004), gr-qc/0406041

\bibitem{MyersPerry}
R. C. Myers, M. J. Perry, \emph{Ann. Phys.} {\bf 172} 304 (1986)

\bibitem{EmparanReall}
R. Emparan, H. S. Reall, A Rotating Black Ring in Five Dimensions,
\emph{Phys. Rev. Lett.} {\bf 88} 101101 (2002), hep-th/0110260

\bibitem{MI}
Y. Morisawa, I. Daisuke, A boundary value problem for the
five--dimensional stationary rotating black holes,
\emph{Phys. Rev.} {\bf D69} 124005 (2004), gr-qc/0401100

\bibitem{kijowski}
E. Czuchry, J. Jezierski, J. Kijowski,  Local approach to thermodynamics
of black holes, \emph{Relativity Today (Proc. 7th Hungarian Relativity
Workshop, 2003)}, Ed. I. Racz, (Akad\'emiai Kiad\'o, Budapest 2004),
gr-qc/0405073

\bibitem{Wald} B. Wald, Black hole entropy is the Noether charge,
\emph{Phys. Rev.} {\bf D48} R3427-R3431 (1993)

\bibitem{acs} A. Ashtekar, A. Corichi, D. Sudarsky,
Hairy Black Holes, Horizon Mass and Solitons, \emph{Class. Quant.
Grav.} {\bf 18}  919-940 (2001)

\bibitem{ashtekar-baez-krasnov}
A. Ashtekar, J. Baez, K. Krasnov,
Quantum Geometry of Isolated Horizons and Black Hole Entropy,
\emph{Adv. Theor. Math. Phys.} {\bf 4} 1-94 (2000), gr-qc/0005126

\bibitem{AK}
A. Ashtekar, B. Krishnan, Isolated and Dynamical Horizons and Their
Applications, gr-qc/0407042 (2004)

\bibitem{CJK2} E. Czuchry, J. Jezierski, J. Kijowski, Dynamics of 
gravitational field within a wave front and thermodynamics of black holes,
\emph{Phys. Rev.} {\bf D70} 124010 (2004), gr-qc/0412042 

\end{thebibliography}
\end{document}